# Sharp Changes of Solar Wind Ion Flux and Density Within and Outside Current Sheets

O. Khabarova (1, 2) · G. Zastenker (1)

*1. Space Plasma Physics Department, Space Research Institute (IKI) of Russian Academy of Sciences, 84/32 Profsoyuznaya Street, Moscow 117997, Russia*

*2. Heliophysical Laboratory, Institute of Terrestrial Magnetism, Ionosphere and Radio Wave Propagation RAS (IZMIRAN), Troitsk, Moscow Region, 142190 Russia*

Phone: +74953331388

Fax: +74953331248

e-mail: olik3110@aol.com `

**Abstract** Analysis of the *Interball-1* spacecraft data (1995–2000) has shown that the solar wind ion flux sometimes increases or decreases abruptly by more than 20% over a time period of several seconds or minutes. Typically, the amplitude of such sharp changes in the solar wind ion flux (SCIFs) is larger than $0.5 \times 10^8$ cm$^{-2}$ s$^{-1}$. These sudden changes of the ion flux were also observed by the Solar Wind Experiment (SWE), on board the *WIND* spacecraft, as the solar wind density increases and decreases with negligible changes in the solar wind velocity. SCIFs occur irregularly at 1 AU, when plasma flows with specific properties come to the Earth's orbit. SCIFs are usually observed in slow, turbulent solar wind with increased density and interplanetary magnetic field strength. The number of times SCIFs occur during a day is simulated using the solar wind density, magnetic field, and their standard deviations as input parameters for a period of 5 years. A correlation coefficient of ~0.7 is obtained between the modelled and the experimental data. It is found that SCIFs are not associated with coronal mass ejections (CMEs), corotating interaction regions (CIRs), or interplanetary shocks; however, 85% of the sector boundaries are surrounded by SCIFs. The properties of the solar wind plasma for days with 5 or more SCIF observations are the same as those of the solar wind plasma at the sector boundaries. One possible explanation for the occurrence of SCIFs (near sector boundaries) is magnetic reconnection at the heliospheric current sheet or local current sheets. Other probable causes of SCIFs (inside sectors) are turbulent processes in the slow solar wind and at the crossings of flux tubes.

**Abbreviations**   SCIF: sharp change of ion flux; IMF: interplanetary magnetic field; SBC: sector boundary crossing; HCS:

## 1. Introduction

Experiments during the space era clearly show that solar wind properties essentially differ at different time and spatial scales (see Marsch and Liu, 1993; Velli and Grappin, 1993). Phenomena with characteristic times ranging from hours to days and even years have been carefully studied for tens of years thanks to regular spacecraft measurements of the interplanetary magnetic field (IMF) and plasma parameters such as the solar wind speed and the density. However, there is a whole class of poorly investigated phenomena, analyzable only based on rather high time-resolution data.

Unique possibilities for studying solar wind small-scale structures appeared in 1995 when the *Interball-1* spacecraft began to measure ion flux $nV$ (where $n$ and $V$ are the ion density and speed respectively) using the Omnidirectional Plasma Sensor (VDP), which had a very high time resolution - no less than 1 s, (for some days it was 60 ms), (Safrankova *et al.*, 1997). The orbit of *Interball-1* allowed the solar wind to be observed during 8 months per year between 1995 and 2000.

One of the results of the *Interball-1* mission was the observation of more than 20 000 sharp borders (characteristic width: $\sim 10^3$–$10^4$ km) of medium-scale solar wind structures (size: $\sim 10^5$–$10^6$ km). The leading and trailing sides of these structures were observed as fast and considerable changes in the solar wind dynamic pressure where the solar wind ion flux abruptly increased or decreased by more than 20% of its initial value within 10 min. Sometimes the ion flux changed several times within seconds.

Small events (amplitude: 0.5–1.0×$10^8$ cm$^{-2}$ s$^{-1}$) were registered near the Earth's orbit, 50 times per day on average, while moderate and sharp ion flux changes (amplitude $\geq 2\times10^8$ cm$^{-2}$ s$^{-1}$) were detected 9 times per day. A list of SCIFs - Sharp Changes of Ion Flux events (when the flux increased or dropped by > 20% within 10 minutes and had an amplitude $\geq 0.5\times10^8$ cm$^{-2}$ s$^{-1}$) was built for 1996-2000 by Riazantseva, Dalin, and Zastenker (2002). Explanation of the SCIFs database creation technique, as well as the results of the investigations of SCIFs' fronts properties, have been published by a group of researchers from the Solar Wind Dynamic Laboratory (IKI) since 2002 (Riazantseva, Dalin, and Zastenker, 2002; Dalin *et al*., 2002a and 2002b; Riazantseva *et al*., 2003a and 2003b; Riazantseva, Khabarova, and Zastenker, 2005; Riazantseva *et al*., 2005; Riazantseva *et al*., 2007).

Since *Interball-1* did not measure the solar wind density and velocity separately, its data were compared with Solar Wind Experiment (SWE) 3 s data, on board *Wind* satellite. Riazantseva, Khabarova, and Zastenker (2005); Riazantseva *et al*. (2005 and 2007) showed that all strong changes in the ion flux with amplitudes $\geq 4\times10^8$ cm$^{-2}$ s$^{-1}$, detected by *Interball-1*, could be found in the *WIND* data as changes in the solar wind density. This is also true for practically all moderate SCIFs with amplitudes $\geq 2\times10^8$ cm$^{-2}$ s$^{-1}$. Thus, when we refer to SCIFs below, we are primarily referring to density changes.

SCIFs are associated with neither interplanetary shock waves nor the boundaries of structures such as magnetic clouds and corotating regions (Riazantseva, Khabarova, and Zastenker, 2005; Riazantseva *et al*., 2005 and 2007). The basic difference between SCIFs and the interplanetary shock waves is the absence of significant changes in the solar wind velocity (Riazantseva, Khabarova, and Zastenker, 2005; Riazantseva *et al*., 2005 and 2007). SCIFs mainly represent large increases or decreases in the solar wind density and resemble compressive fluctuations, which have been known since 1990 (Bruno and Carbone, 2005). However, the typical timescales for these phenomena are different (hours for compression fluctuations and minutes or even seconds for SCIFs).

Preliminary investigations have shown that SCIFs are surrounded by rather slow, but dense solar wind (Riazantseva, Khabarova, and Zastenker, 2005; Riazantseva *et al*., 2007). The other important property of SCIFs is their geoefficiency. The influence of SCIF-caused sharp impulses of the solar wind dynamic pressure on the terrestrial magnetosphere causes significant geomagnetic field changes, local aurora borealis enhancements, and excitation of geomagnetic pulsations of different types in different geomagnetic latitudes (Borodkova *et al*., 2005; Parkhomov, Riazantseva, and Zastenker, 2005).

While the properties of small-scale solar wind structures, such as SCIFs, have been investigated in depth, we still do no know their origin. The following questions have yet to be answered:

1. Do SCIFs occur as a result of stochastic processes in the solar wind? Alternatively, does the frequency of SCIFs' occurrence at 1 AU depend on the properties of the solar wind surrounding SCIFs?

2. Are the studied SCIFs consequences of processes on the Sun (i.e. are they related to the solar structures, keeping their form and properties while propagating from the Sun to the Earth)? Alternatively, do SCIFs occur directly in the solar wind plasma as the result of the processes taking place in space (i.e. turbulence or instabilities in the solar wind plasma)?

3. What is the lifespan of SCIFs?

For the best understanding of the processes observed in near-Earth space, we must investigate the properties of medium- and large-scale SCIFs. Here, we study the first question in detail and make assumptions on the nature of SCIFs observed at 1 AU. The analysis of solar wind plasma conditions related to SCIFs includes a case study, statistical analysis of experimental data, and modelling.

## 2. Sharp density changes occurring at 1 AU and the corresponding solar wind conditions

### 2.1. A CASE STUDY

A typical case of a sharp solar wind density or ion flux change near 1 AU is shown in Figure 1. Figure 1a shows the SCIFs on 26 April, 1998, which were measured by both the Hot Plasma and Charge Particles (3DP) instrument, on board *Wind*, (time resolution 3 s) and *Interball-1* VDP instrument (time resolution 1 s), with a time delay of ~1.5 h, as spacecraft were at a distance of ~200 Re one from other. For illustration purposes, the data from *WIND* are time shifted to match the *Interball-1* data in Figure 1a. We use arrows to indicate the start times of SCIFs (intense increases or decreases in ion flux with amplitudes $\geq 2\times10^8$ cm$^{-2}$ s$^{-1}$) on the *Interball-1* ion flux curve.

Despite a slight transformation along the propagation path of the streams containing SCIFs, it is easy to observe similar sharp changes in the solar wind density (measured by *WIND*) and ion flux (measured by *Interball-1*). More examples of SCIFs measured by *Interball-1* and their corresponding density changes (measured by the *WIND* and *IMP8* spacecraft) can be found in papers by Riazantseva, Dalin, and Zastenker (2002), Riazantseva *et al*. (2003a and 2003b),

Riazantseva, Khabarova, and Zastenker (2005), Riazantseva *et al*. (2005), and Dalin *et al*. (2002b).

Figure 1a shows that small-scale boundaries of medium-scale flows are rather stable and do not disappear during solar wind propagation at ~200 $R_e$. Thus, SCIFs are not a result of small-scale instabilities (in the opposite case their life-time would be significantly shorter) and they are not specific features of the Earth magnetosphere foreshock region (as otherwise they would be observed only by *Interball-1*). Moreover, according to Dalin *et al*. (2002a and 2002b), there are examples of SCIFs that remained stable for distances up to 0.6 AU. Therefore, either SCIFs originate at the Sun with their consequent transport by solar wind streams, or they are a result of some large-scale processes in space.

The OMNI2 time series of hourly averaged solar wind parameters are given in Figures 1b, c for the entire day of 26 April, 1998. Vertical boxes in Figure 1b show the number of SCIFs per hour with amplitudes $\geq 0.5 \times 10^8$ cm$^{-2}$ s$^{-1}$. If we analyse the properties of the streams that carried SCIFs to the Earth's orbit, we see that the substantial growth in the number of SCIFs' per hour is accompanied by a significant growth in the solar wind density and its standard deviation (Figure 1b), while other key background parameters remain stable (Figure 1c).

Visual analysis of *Interball-1* and OMNI2 data has shown that most days with high SCIF number are characterized by plasma conditions similar to those represented in Figure 1. We will confirm this statement using statistical analysis below. Thus, SCIFs are not a result of random processes in space plasma, but are structures related to streams with specific conditions.

One confirmation of the previous idea is the way SCIFs are grouped: days with high number of SCIFs alternate with the days without SCIFs or with a very small number of SCIFs. Nine events per day (as was mentioned in the Introduction) is an average rate of the occurrence of SCIFs at the Earth's orbit. It does not reflect the actual SCIFs observation frequency; therefore, we have to carry out a more careful investigation of the temporal distribution of SCIFs.

## 2.2. FEATURES OF SCIFs TEMPORAL DISTRIBUTION

In this section, we analyse the statistical properties of 5300 SCIFs, with amplitudes larger than $2 \times 10^8$ cm$^{-2}$ s$^{-1}$, observed on 427 of the 673 days when *Interball-1* was in the solar wind (from 28 February, 1996 to 21 September, 2000). To ensure clean results, all days where *Interball-1* passed through the foreshock region were not included.

The time distribution of the number of SCIFs per day ($N_{SCIF}$) is shown in Figure 2. The horizontal axis represents the daily number of SCIFs, and the vertical axis represents the class frequency (the number of data points that fall inside the class interval) in percentage of the entire number of SCIFs. This histogram is drawn as follows: we take the number of SCIFs per day ($N_{SCIF}$) and multiple it by the number of days ($N_{days}$) when the given $N_{SCIF}$ is observed. For example, 4 SCIFs per day were observed by *Interball-1* 34 times during the 1996–2000 observations. Hence, whole number of SCIFs, observed with such frequency is 136. Then, we divide the x-axis into several intervals such as $0 \leq N_{SCIF} < 2$ and $2 \leq N_{SCIF} < 4$ and calculate the number of SCIFs occurring in the specific range of $N_{SCIF}$ values.

The obtained distribution is significantly shifted from a Gaussian: about 50% of the total number of events were observed from 17 to 64 times per day. This demonstrates the grouping effect; on some days, SCIFs are observed sequentially in a pulse packet that probably contains the small-scale boundaries of some medium- or large-scale solar wind structures.

The results of this statistical analysis allow us to assume that it is possible to evaluate $N_{SCIF}$ as a function of parameters of the ambient solar wind. We will look for the most characteristic changes of key solar wind parameters during the periods with large numbers of SCIFs and build a modelling function on the basis of the parameters best correlated with $N_{SCIF}$.

## 2.3. BEHAVIOUR OF SOLAR WIND PARAMETERS DURING PERIODS OF OBSERVATION OF SCIF PACKETS AT 1 AU

### 2.3.1 Analysis of histograms

The characteristics of the streams containing SCIFs will be analysed taking into account the statistical properties of days with high number of SCIFs. We consider a day to be a 'SCIF pulse-packet day' if the number of SCIFs exceeds five that day ($N_{SCIF} \geq 5$). There were 264 days (containing 4 951 SCIFs) that satisfied this condition. This means that 93% of SCIFs were observed during 'SCIFs pulse-packet days' (which makes up 62% of the total number of days); therefore, the grouping effect is strong.

Analysis of the solar wind plasma properties for those days shows that SCIFs-containing large and medium-scale structures can be characterised by enhanced solar wind density $n$ (Figure 3a), slightly increased average IMF magnitude |**B**| (Figure 3b), and an increased standard deviation of both (Figures 3c and d). The white boxes in Figure 3 represent the distribution of parameters for the days with high number of SCIFs for the period 1996–2000, and the black boxes show the distribution of the same parameters for the same time range according to *WIND* daily data. The standard deviations shown in Figures 3c and 3d are calculated using the hourly *WIND* data. In all cases, there is a shift of the white histograms to the right, relative the black ones, especially for the density and its standard deviation. The statistical characteristics of the histograms are listed in Table 1, where the distributions for the entire time period of measurements and for days when $N_{SCIF} \geq 5$ are marked as 'all' and 'scif', respectively.

According to a t-test, the difference between all pairs of 'all-scif' variables is statistically significant (the t-test has a conventional significance level less than 0.05 ($p < 10^{-6}$). This means that the histogram shifts in Figure 3 are not obtained by chance.

Another interesting feature of the 'scif' histograms is that their skewness values are lower than the corresponding ones for 'all' data (see Table 1). Skewness measures the deviation of the distribution from symmetry. If the skewness is clearly different from 0, then the analyzed distribution is asymmetrical, while normal distributions are perfectly symmetrical. The asymmetry of the solar wind parameters' distributions is an evidence of some structuring of the solar wind plasma. In our case, the closeness of 'scif' distributions to a Gaussian could mean that stochastic processes more often occur in the plasma containing SCIFs or that such plasma streams originate far from the solar wind source.

There is one more confirmation that SCIFs are observed in the turbulent solar wind. We compared the IMF variability in the ultra low frequency (ULF) band (ULF wave index) for days with high SCIF number to the variability over the entire time period of observations (Figure 4). The ULF index is a 1-h resolution index, which characterises the turbulence level of the solar wind magnetic field in the ULF range (Romanova *et al*., 2007) and is calculated from three components of the IMF (measured by *WIND* or *ACE* spacecraft) with 1 min resolution. The higher the ULF-index value, the higher is the IMF disturbance level in the 1–10 mHz frequency range.

The black histogram shown in Figure 4 represents the distribution of the ULF index for 1996–2000, and the white histogram shows the distribution of the ULF index for the days when the number of SCIFs per day observed by *Interball-1* is larger or equal to five. The shift of the white histogram to the right denotes a high level of magnetic field turbulence in the solar wind streams containing SCIFs.

It is interesting to note that the difference between the histograms in Figures 3 and 4 remains statistically significant for amplitudes lower than $2 \times 10^8$ cm$^{-2}$ s$^{-1}$, but the higher the SCIF amplitudes the more significant are the characteristic shifts. Similar properties have been recently found by Riazantseva, Khabarova, and Zastenker, 2005; Riazantseva *et al*., 2007, as a result of the analysis of the histograms of solar wind parameters computed in a range of 30 min around the observation of SCIFs. This demonstrates the existence of medium- or large-scale dense, turbulent regions, carrying SCIFs to 1 AU.

*2.3.2. A superposed epoch analysis*

A method of superposed epoch analysis is often applied to time series in solar-terrestrial physics to study the conditions accompanying repeated events (see, for example, Lavraud *et al*., 2005). The main concept of the superposed epoch analysis method is that data averaging due to superposition of several curves purifies the useful signal and suppresses the noise. If the effect is absent, the analysis result appears as a stochastic curve (or even as a straight line). On the contrary, statistically significant results are obtained when the extreme points with their standard deviations are beyond the 95% confidence interval, plotted on each side of the mean value line.

Figure 5 shows the behaviour of the main plasma parameters averaged over the days with high SCIFs number. Day zero corresponds to the day of observation with 5 or more SCIFs (264 cases). We have put all the statistical information in Table 2 to clearly show the effect.

The increases in density, interplanetary magnetic field, and their standard deviations for days with high SCIFs number are confirmed by the superposed epoch analysis results. Figures 5a and b show that a significant increase in the parameters is observed in the range of two days around day zero.

The behaviour of the geomagnetic Kp index is an indirect confirmation of the geoefficiency of SCIFs (or of streams containing SCIFs). The Kp index increases during the day, when the SCIF packet interacts with the terrestrial magnetosphere (see Figure 5c). The effect lasts for two days. The interesting fact is that the solar wind speed decreases before day zero and increases symmetrically after that, though the changes are rather small.

### 2.4. LINEAR CORRELATION ANALYSIS

We performed a correlation analysis of the solar wind key parameters with frequency of SCIFs at 1 AU. It was found that the number of SCIFs per day does not correlate with the OMNI2 time series of the daily averaged solar wind speed *V*, the electric field **E** = -*V*×**Bz**, plasma beta parameter β, or the Alfven Mach number, and it poorly correlates with the standard deviation of *V*. The correlation coefficients between $N_{SCIF}$ and these parameters do not exceed 0.22, as shown in Table 3.

We have removed from the σ*n* time series the hour time intervals where SCIFs occurred to avoid that they increase the standard deviation of the density and to ensure the absence of artefacts in our statistical analysis and modelling. Results of the correlation analysis between the number of SCIFs per day and solar wind parameters, shown in Figure 3, are summarized in Table 4. The $N_{SCIF}$ time series show behaviour similar to that of the density, the IMF averaged magnitude, and their standard deviations. The correlation coefficients listed in Table 4 reach up to 0.5.

Thus, if we want to find a modelling parameter, characterising the frequency of the occurrence of SCIFs at 1 AU as a function of some solar wind parameters, we have to focus on the solar wind density, the interplanetary magnetic field, and their variabilities.

## 3. Modelling

A composite function method has been used for modelling. This method assumes that if the parameters, taken separately, correlate with a variable just moderately, their optimal combination could give a higher correlation with this variable. After the correlation analysis, the positive correlating parameters are placed in the numerator and the negative ones are placed in the denominator. An expert evaluation, in combination with computer coefficient adjustments, gives the best chance to find the optimal parameters for simulating the variable. The method is analogous to the neural network method and demands an extremely good knowledge of simulated processes.

The result of seeking various modelling functions to find the most effective fitting parameter, $P_{SCIF}$, to simulate the number of SCIFs per day, $N_{SCIF}$, includes plasma and magnetic field parameters. This is expressed as follows:

$$P_{SCIF} = -2.398 + 0.0267 \times k_n \times (4 \times n + \sigma n) \times k_B \times (|\mathbf{B}| + 3 \times \sigma|\mathbf{B}|) \qquad (1)$$

where $n$ is the solar wind density [cm$^{-3}$]; $|\mathbf{B}|$ is the IMF averaged magnitude [nT]; $\sigma n$ is the standard deviation of the solar wind density [cm$^{-3}$], with the hours with SCIFs removed to avoid artefacts; $\sigma|\mathbf{B}|$ is the standard deviation of interplanetary magnetic field averaged magnitude [nT]; $k_n = 1$ [cm$^3$], and $k_B = 1$ [nT$^{-1}$] are factors to preserve $P_{SCIF}$ dimensionless.

As we can see from (1), there are two multipliers, which represent density and the IMF input. Synchronous increasing of density and IMF together with their variations provides the best conditions for SCIFs origination or propagation. As it will be shown below, such solar wind conditions exist around current sheets (sector boundaries).

An example of the modelling for 2000 is shown in Figure 6, where the observed frequency of SCIFs by *Interball-1* is shown in comparison with the fitting parameter $P_{SCIF}$ (Figure 6a), its multipliers $4 \times n + \sigma n$ (Figure 6b), and $|\mathbf{B}| + 3 \times \sigma|\mathbf{B}|$ (Figure 6c). A rather good match of the $P_{SCIF}$ parameter with the observed data is found.

It is remarkable that the correlation coefficients between the observed number of SCIFs per day and all the parameters, included in $P_{SCIF}$, are no more than 0.5 (see Table 4 and Figures 6b and c). At the same time, the correlation coefficient between $N_{SCIF}$ and the modelling function $P_{SCIF}$ is 0.7. All these correlation coefficients are calculated for the entire period of observations (1996–2000).

$P_{SCIF}$ and $N_{SCIF}$ have identical means and close standard deviations (8.8 for $P_{SCIF}$ and 13.0 for $N_{SCIF}$). Therefore, from a statistical point of view, they follow very similar trends and do not coincide by chance.

These facts confirm the success of the simulation and let us conclude that the IMF, solar wind density and their variabilities, contribute to the stability and propagation (or even occurrence) of SCIFs.

## 4. Current sheets and sharp changes of solar wind density and ion flux

Now, it is reasonable to consider the physical meaning of the discussed phenomenon. Rather important facts can be found by comparing the time when the SCIFs were measured with the arrival of structures such as magnetic clouds (MCs), corotating interaction regions (CIRs), and sector boundaries. The dates of the beginning and end of MCs and CIRs passages, as well as the sector boundary crossings (SBCs) for 1996–2000, were taken from an open source catalogue (International Solar Terrestrial Physics (ISTP) Solar Wind Catalogue of Candidate Events). A SCIF event was considered to be associated with one of these large-scale structures if it occurred within a time interval starting a day before the structure arrival at 1 AU and ending a day after its termination.

It was found that SCIFs are practically not associated with the first two structures, i.e. no more than 2% of them are located within (or around) MCs or CIRs (Khabarova and Zastenker, 2008). On the other hand, an overwhelming majority of sector boundaries in 1996–2000 (85%) were surrounded by SCIFs.

The analysis has shown no increase or decrease in the SCIF amplitudes within the sector boundary areas. We simply observe a stable increase in the SCIF number in the sector boundary vicinity. The analysis shows that 38% of all cases of SCIF observations having between 9 and 64 events per day correspond to the current sheet crossings. Meanwhile, 64 SCIFs per day are observed near sector boundaries only in 25% of the cases.

It is remarkable that the features of solar wind plasma for 141 non-SBC-related days with high SCIF number, analysed in the sections 2.3.1 and 2.3.2, are the same as those for 123 SBC-related ones. Therefore, although we cannot explain the occurrence of all SCIFs at the Earth's orbit by processes in only the current sheets, analysis of solar wind conditions at currents sheets and around them probably could give us a clue to understand the nature of the entire body of events.

## 4.1 BEHAVIOUR OF SOLAR WIND PARAMETERS IN CURRENT SHEETS

An increase in the ion flux near sector boundaries was first mentioned in 1984 by Briggs and Armstrong, 1984; however, the nature of this phenomenon has not practically been investigated after this work. Let us look closely at the plasma properties in the sector boundary vicinities and compare them with the typical plasma characteristics where SCIFs were observed.

The commonly accepted picture describes sector boundaries as a result of the intersection of the heliospheric current sheet (HCS), which is formed by an extension of the main solar neutral line (the heliomagnetic equator). The HCS, discovered by Wilcox and Ness, 1965, divides the heliosphere into the areas of opposite IMF direction called sectors (Svalgaard *et al*., 1975). It is known that sector boundary crossings (SBCs) are accompanied by an increase in the plasma density and a decrease in the solar wind speed near zero-line – the line of the IMF inversion (Schwenn, 1990; Smith, 2001; Crooker *et al*., 2004; Blanco *et al*., 2006).

Theoretically, a sector boundary must be associated to the HCS; however, in reality, this view is simplistic and only appropriate for educational purposes. Detailed investigations show that problems arise with the identification of the HCS based on SBC data. First, the main solar zero-line does not always coincide with sector boundaries due to the complexity of the magnetic field when going from the Sun to the Earth, as shown by Crooker *et al*., 2004. Loop structures can entangle the IMF and form local current sheets with field reversals, which are not true HCS sector boundaries. The local zero-lines between groups of sunspots, when reaching 1 AU, can be also identified as current sheets (or sector boundary crossings).

Another problem is that the HCS and even the local current sheets are not truly sheets. Due to their complex internal structure or undulating motion, the crossing of these large-scale structures at 1 AU can last for several hours or even days. The result is that the spacecraft meets several sector boundaries for one current sheet (Blanco *et al*., 2006). This fact complicates the identification of sector boundary crossings and leads to contradictory results, as investigators sometimes consider the heliospheric current sheet to be a very thin layer and believe that its crossing takes no more than several minutes.

Figure 7 shows the distribution of SBC durations after the ISTP Solar Wind Catalogue of Candidate Events for 1994–2000. Change of the IMF direction (for example, from Sun-outward to Sun-inward) generally takes no more than one day (149 cases), as illustrated in Figure 7. Two-day length SBCs were observed 49 times only, with very few cases corresponding to 3 to 8 days of unstable IMF direction.

The methods to identify sector boundaries are slightly different. SBCs are mainly determined by changes in the interplanetary magnetic field longitude angle $\phi_B$ (IMF azimuth) due to changes in the horizontal Bx and By IMF components. Sometimes researchers simply look at the change of sign of the IMF Bx component, and sometimes they additionally use suprathermal electron data to build lists of SBC dates (see Crooker *et al.*, 2004). Geomagnetic field data were also used for this purpose, especially before the space age (Svalgaard, 1975). All this may lead to inconsistent results and misunderstandings in the analysis of the properties of the HCS.

Consequently, there are several public lists of SBC dates that agree poorly (see, for example, ISTP Solar Wind Catalogue http://www-spof.gsfc.nasa.gov/scripts/sw-cat/grep-ls/sw-cat-categories.html; The Wilcox Solar Observatory List http://wso.stanford.edu/SB/SB.html, and the OMNIweb list http://omniweb.gsfc.nasa.gov/html/polarity/polarity_tab.html) and some 'private' lists, made by several researchers for their own scientific purposes (see, for example, Leif Svalgaard's list at http://www.leif.org/research).

The behaviour of solar wind parameters in the HCS and local current sheets has been analysed to compare their key properties with features of SCIF packets (see Figures 5 and 8). To obtain the best statistics and avoid mistakes, we used a method of superposed epoch analysis for two of SBC lists: the list by Leif Svalgaard, containing 1 300 events for the period of available OMNI2 data from January, 1964 to April, 2010 (Figures 8a, b, and c) and the ISTP Solar Wind Catalogue Candidate Events for the above mentioned 149 one-day sector boundary crossings between 1994 and 2000 (see Figures 8d, e, and f).

The typical profiles of solar wind parameters along the day of sector boundary crossing are shown in Figure 8, and their statistical properties are listed in Table 5 (left panel of Figure 8) and Table 6 (right panel of Figure 8).

The IMF strength growths on the SBC days, as we can see from Figures 8a and d. But, from a theoretical point of view, the average IMF module must drop at the time of crossing. There is no contradiction in this. Precise data show that all the IMF components drop at the current sheet for very short time (several minutes) and that the current sheet is surrounded by regions of increased IMF. Therefore, the IMF increase at a day zero is the result of 24-h averaging.

The well-known growth of the solar wind density across the heliospheric current sheet is observed in a wide time range, from one day before the SBC to two days after (Figures 8b and e).

The increased variabilities of density and of the IMF strength around sector boundaries paint a complicated picture of instabilities, which develop at the current sheets. As a result, the current sheet plasma tends to be highly disturbed.

Growth of geomagnetic activity at the heliospheric current sheet crossing, represented by the Kp index (see Figures 8c and f), is an interesting feature, which has been investigated for many years (see the pioneer works by Hirshberg and Colburn, 1973; Hakamada, 1980).

The decrease of the Kp index one or two days before a sector boundary crossing is not a well-known effect, in spite of its discovery in 1973 by Leif Svalgaard (Svalgaard, 1973). This phenomenon was recently revisited by Watari and Watanabe (2006), who investigated typical change of the Kp index profile (like in Figures 8c and f, but based on the OMNIweb SBC list) with the solar cycle.

The solar wind speed decrease before a sector boundary and its subsequent increase is a rarely discussed phenomenon, although this effect was also mentioned in the pioneer works by Svalgaard (1973 and 1975). Usually, the speed is considered to be lower around the HCS (Borrini *et al.*, 1981). The nature of the non-symmetric profiles of Kp and the solar wind speed in the vicinity of current sheets has been investigated only in a few works. In a case study Neugebauer *et al.* (2004) mentioned the decreasing speed of the solar wind before the observed sector boundaries and after them. Referring to von Steiger *et al.* (2000), the authors believe that it is possible to explain this phenomenon because of the various characteristics of streams before and after the HCS. Lacombe *et al.* (2000) suppose that the solar wind speed depression before the HCS is a feature of the high-pressure solar wind, which is a result of dynamical stream interactions.

The interesting and intriguing fact is the almost full similarity of the behaviour of the parameters in Figures 8 and 5. This means that either typical conditions inside and nearby HCS are ideally suited for maintain and transference of SCIFs in the solar wind plasma, or that the HCS is a place where they originate.

## 4.2 THE PHYSICAL NATURE OF THE OBSERVED SHARP DENSITY CHANGES WITHIN CURRENT SHEETS AND IN THE SLOW SOLAR WIND

We showed in 2.3 that SCIFs are observed in dense and turbulent regions of the solar wind. We also found out in previous studies that the speed of the solar wind surrounding SCIFs is lower than usual (Riazantseva *et al*., 2007). The increased turbulence in dense, slow plasma leading to large-scale instabilities at the HCS or the local current sheets could be a cause of SCIF occurrence at 1 AU. Properties of near-HCS zone of increased turbulence, containing discontinuities, are still insufficiently investigated (Crooker *et al*., 2004; Blanco *et al*., 2006; Marsch, 2006). Roberts, Keiter, and Goldstein (2005) noticed that many dynamic processes go on permanently inside the HCS and that its structure becomes more and more turbulent and complex with heliocentric distance.

Any large-scale instability near the IMF zero-line inside the HCS can be a cause of magnetic reconnection. The results of many authors confirm this idea, which is obvious from the general reasons (for example, Murphy *et al*., 1993; Gosling *et al*., 2006; Phan, Gosling, and Davis, 2009).

The heliospheric current sheet not only extends along its propagation from the Sun, but also is enriched by repeated reconnections at the zero-line. Waves, discontinuities, and soliton-like structures are observed by many spacecraft both at sector boundaries and in their near vicinity. Probably, discussed SCIFs are beamlet-structures (double ion beams), which sometimes are observed in the vicinity of sector boundaries. Hammond *et al*. (1995) described them and postulated that these beams are a result of magnetic reconnection.

Since the solar wind properties for days with SBC-related SCIFs and non-SBC-related ones are the same, the question now arises as to whether SCIFs', which are observed far from sector boundaries, and HCS-associated SCIFs, have the same nature?

Neugebauer *et al*. (2004) noticed that the non-HCS slow solar wind includes a lot of small-scale structures, such as discontinuities, magnetic holes, and low-entropy structures. These structures are usually associated with the slow solar wind around the HCS. Therefore, turbulent processes in the slow solar wind could be a key cause of SCIFs observed in the non-HSC solar wind.

We can assume that the non-HCS SCIFs can be explained by the presence of flux tubes in the solar wind. The concept of separated thin plasma tubes (or spaghetti-like structures) existing in the solar wind plasma has been put forward repeatedly for more than 40 years (see the reviews by Wang and Sheeley, 1990; Hollweg, 1972 and 1986; Wang, 1993; Li, 2003) since Parker (1963) first suggested it. Recently, Borovsky (2008) presented a rather convincing evidence for the existence of plasma tubes and reached the important conclusion that the tubes are larger in the slow wind than in the fast wind. It has been estimated that the "the median size of the flux tubes at 1 AU is $4.4 \times 10^5$ km".

SCIFs could be a sign of crossings of such tubes, as SCIFs in the SCIF-packet are usually observed for several minutes, sometimes hours (see an example in Figure 1a, where SCIFs are observed more frequently than 1 h from 8 h to 15 h and more rarely from 17h to 21h). This corresponds to a distance of about $10^5$–$10^6$ km (Riazantseva, Dalin, and Zastenker, 2002; Dalin *et al*., 2002a and 2002b; Riazantseva *et al*., 2003a and 2003b). Therefore, the characteristic size of the structures with sharp borders (detected as SCIFs) coincides with the estimated sizes of the flux tubes.

Qin and Li (2008) have recently developed a model of the solar wind turbulence, which consists of independently moving flux-tube structures (cells). They believe that local current sheets (not the HCS) are possibly the boundaries of such individual flux tubes.

Thus, both magnetic reconnection and turbulence are possible causes of non-HCS SCIFs in the slow solar wind. Another possibility is that such SCIFs are the result of reconnection directly on the Sun in large coronal loops, which form slow solar wind streams (von Steiger *et al*., 2000).

## 5. Conclusions

The sharp change of the solar wind ion flux, SCIF, is a very fast and abrupt process. SCIFs were found during the analysis of the *Interball-1* spacecraft high-resolution data (1995–2000). They take from seconds to minutes to cross the spacecraft, and the solar wind ion flux can increase/decrease several times during these passages. After comparing the *Interball-1* SCIF database with *WIND* data, it was found that SCIFs are primarily changes of density.

SCIFs are not a feature of the foreshock area ahead the Earth magnetosphere. They are rather long-living structures, sometimes they are traced up to a distance of 0.6 AU. SCIFs are not a result of local instabilities, but they are related to some large-scale processes in the solar wind (or, possibly, on the Sun). SCIFs are not associated with interplanetary shocks, CIRs or CMEs, but frequently observed near sector boundaries.

On the basis of the current investigation, we can conclude that

1. SCIFs with amplitudes larger than $2\times10^8$ cm$^{-2}$ s$^{-1}$ usually arrive to the Earth's orbit in a pulse packet, with up to 128 events per day (we define that a pulse-packet is observed when number of SCIFs per day exceeds five). 93% of SCIFs measured by *Interball-1* belonged to pulse packets, which were observed during 62% of all days. The grouping effect is proved both by a case study and by statistical analysis. This can be a sign of the episodic occurrence of solar wind streams containing discontinuities at 1 AU.

2. Analysis of the solar wind properties of streams containing SCIFs shows that SCIFs are observed under dense, turbulent solar wind conditions, with slightly increased values of the IMF strength. For example, a superposed epoch analysis shows that the standard deviations of density and IMF are correspondingly 1.9 and 1.6 times larger in days with high SCIF number than in the days with no SCIFs. The density increases by a factor of 1.4, and the IMF magnitude increases by a factor of 1.2 during the days where SCIF pulse-packets are observed. In combination with the results of previous work that shows that SCIFs mainly occur in the slow solar wind, this lets us conclude that SCIFs hardly related to active solar processes such as solar flares and CMEs. They are also not generated in (or around) dense solar wind regions like CIRs, but they could be a result of turbulence in the slow solar wind or in the solar corona.

3. The number of SCIFs per day can be successfully simulated by a combination of key solar wind parameters comprising the solar wind density, interplanetary magnetic field, and their variabilities. The correlation coefficient between the modelling parameter and the observed number of SCIFs per day is 0.7. This means that the occurrence of SCIFs at the Earth's orbit is not a random process, but a result of specific plasma conditions. As we can see from the previous conclusion, Streams, containing SCIFs, have properties that are significantly different from the properties of the ambient solar wind. Simulations show that SCIFs most frequently occur when both the solar wind density and IMF strength increase (as well as their variability). At 1 AU, such plasma properties are mainly observed at sector boundaries in the solar wind.

4. SCIFs are associated with sector boundary crossings. 85% of all tested sector boundaries were surrounded by SCIFs. The conditions favourable for the origin and propagation of SCIFs exist in the ±1 day vicinity of the HCS and local current sheets.

5. A considerable percentage of SCIFs (60% from the total) are observed far from the sector boundaries (inside outward or inward sectors) in the slow solar wind when HCS-like conditions exist. On the basis of the obtained results, we hypothesize that HCS-like conditions play a key role in formation of the discussed ion flux (density) changes.

As the typical time span between SCIFs is several (or tens) minutes and their passage time is no more than 10 min, we can consider SCIFs as the borders (with characteristic size $\sim 10^3$–$10^4$ km) of some plasma structures with a width of $\sim 10^5$–$10^6$ km. These plasma structures have specific properties and represent solar wind streams, observed both around sector boundaries and inside sectors.

We believe that the nature of SBC-related SCIFs could be investigated analysing the reconnection process both at the HCS and local current sheets (which are separators of sunspot groups of the opposite sign), as discontinuities around HCS mainly occur as the result of turbulence and repeated magnetic reconnection at a magnetic X-line.

SCIFs, not related to SBC, could indicate crossings of flux tubes in the solar wind. Such tubes are supposed to exist in relatively slow solar wind, undisturbed by fast streams like CMEs. It is remarkable that their average size is estimated as $\sim 10^5$–$10^6$ km, which coincides with the size of the structures containing SCIFs.

The other possible origin of SCIFs is turbulence. The slow solar wind is more turbulent than the fast one. As a result, numerous discontinuities are a typical feature of the slow solar wind.

All the mechanisms listed above will be the subjects of future testing and investigations. The analysis of distant spacecraft data could be particularly useful as it could help to draw the picture of the origin and propagation of SCIFs in the solar wind, and could answer some key questions about their nature.

**Acknowledgements** The one second plasma data are obtained in the result of the Interball mission (see http://www.iki.rssi.ru/interball). *WIND* 3DP solar wind data with 3 second resolution are from web-site of Goddard Space Flight Center: http://cdaweb.gsfc.nasa.gov , and OMNI2 data are taken from the official OMNIweb site: http://omniweb.gsfc.nasa.gov/ow.html. The detailed list of sector boundaries crossings (ISTP Solar Wind Catalogue of Candidate Events) is presented on http://www-spof.gsfc.nasa.gov/scripts/sw-cat/grep-ls/SBC.html. SBC List by Leif Svalgaard was taken from his official web-page: http://www.leif.org/research/sblist.txt. Data, articles and the detailed description of ULF-index calculation technique can be found on ftp://space.augsburg.edu/maccs/ULF_Index. The authors wish to thank Dr. Elena Driver (St.Mary's University College, London) for her linguistic remarks. This research was supported by RFBR's grant 10-02-00277-a, and partially by 10-02-01063-a grant.

# References

Blanco, J.J., Rodriguez-Pacheco, J., Hidalgo, M.A., and Sequeiros, J.: 2006, Analysis of the heliospheric current sheet fine structure: Single or multiple current sheets. *J. Atm. Sol.-Ter. Phys*. **68**, 2173–2181.

Borodkova, N.L., Zastenker, G.N., Riazantseva, M.O., and Richardson, J.D: 2005, Large and sharp solar wind dynamic pressure variations as a source of geomagnetic field disturbances in the outer magnetosphere (at the geosynchronous orbits. *Planet. Space Sci*. **53**, 25-32.

Borovsky, J.E.: 2008, Flux tube texture of the solar wind: Strands of the magnetic carpet at 1 AU. *J. Geophys. Res*. **113**, A08110.

Borrini, G., Gosling, J., Bame, S., Feldman, W., and Wilcox, J.: 1981, Solar wind helium and hydrogen structure near the heliospheric current sheet: a signal of coronal streamers at 1 AU. *J. Geophys. Res*. **86**, 4565-4573.

Briggs, P.R., and Armstrong, T.P.: 1984, Observations of interplanetary energetic ion enhancements near magnetic sector boundaries. *Geophys. Res. Lett*. **11**, 27–30.

Bruno, R., and Carbone, V.: 2005, The solar wind as a turbulence laboratory. *Living Rev. Solar Phys.* **2**, http://www.livingreviews.org/lrsp-2005-4.

Crooker, N.U., Huang, C.-L., Lamassa, S.M., Larson, D.E., Kahler, S.W., and Spence, H.E.: 2004, Heliospheric plasma sheets. *J. Geophys. Res*. **109**, A03107.

Dalin, P.A., Zastenker, G.N., Nozdrachev, M.N., and Veselovsky, I.S.: 2002a, Properties of large and sharp impulses in the solar wind. *Int. Journ. Geom. Aeron*. **3**, 51-56.

Dalin, P.A., Zastenker, G.N., Paularena, K.I., and Richardson, J.D.: 2002b, A survey of large, rapid solar wind dynamic pressure changes observed by Interball-1 and IMP 8. *Annales Geophysicae*, 20, 293–299.

Gosling, J.T., McComas, D.J., Skoug, R.M., and Smith, C.W.: 2006, Magnetic Reconnection at the Heliospheric Current Sheet and the Formation of Closed Magnetic Field Lines in the Solar Wind. *Geophys. Res. Lett*. **33**, L17102.

Hakamada, K.: 1980, Geomagnetic activity at the time of heliospheric current sheet crossings, *Geophys. Res. Lett*. **7**, 653–656.

Hammond, C.M.; Feldman, W.C.; Phillips, J.L.; Goldstein, B.E.; Balogh, A.: 1995, Solar wind double ions beams and the heliospheric current sheet. *J. Geophys. Res.* **100**, 7881-7889.

Hirshberg, J. and Colburn, D.S.: 1973, Geomagnetic activity at sector boundaries, *J. Geophys. Res*. **78**, 3952–3957.

Hollweg, J.V.: 1972, Supergranulation-driven Alfvén waves in the solar chromosphere, and related phenomena. *Cosmic Electrodynamics* **2**, 423-444.

Hollweg, J.V.: 1986, Transition region, corona, and solar wind in coronal holes. *J. Geophys. Res.* **91**, 4111-4125.

Khabarova, O.V. and Zastenker, G.N.: 2008, Sharp and sizeable changes of solar wind ion flux as a feature of dense non-CIR turbulent regions, Geoph. Res. Abstracts, 10, EGU2008-A-09908.

Lacombe, C., Salem, C., Mangeney, A., Steinberg, J.-L., Macsimovic, M., and Bosqued, J.M.: 2000, Latitudinal distribution of the solar wind properties in the low- and high-pressure regimes: Wind observations. *Ann.Geoph.* **18**, 852-865.

Lavraud, B., Denton, M.H., Thomsen, M.F., Borovsky, J.E., and Friedel, R.H.W.: 2005, Superposed epoch analysis of dense plasma access to geosynchronous orbit. *Ann. Geophys.* **23**, 2519–2529.

Li, X.: 2003, Transition region, coronal heating and the fast solar wind. *Astron. and Astrophys.* **406**, 345–356.

Marsch, E.: 2006, Kinetic Physics of the Solar Corona and Solar Wind. *Living Rev. Solar Phys.* **3**, 1, http://www.livingreviews.org/lrsp-2006-1.

Marsch, E. and Liu, S.: 1993, Structure functions and intermittency of velocity fluctuations in the inner solar wind. *Ann. Geophys.* **11**, 227-238.

Murphy, N., Smith, E.J., Burton, M.E., Winterhalter, D. and McComas, D.J.: 1993, Energetic ion beams near the heliospheric current sheet: possible evidence for reconnection. *Jet Propulsion Lab.(NASA) Technical Report,* http://trs-new.jpl.nasa.gov/dspace/bitstream/2014/35959/1/93-1689.pdf .

Neugebauer, M., Liewer, P.C., Goldstein, B.E., Zhou, X., Steinberg, J.T.: 2004, Solar wind stream interaction regions without sector boundaries. *J. Geophys. Res.* **109**, A10102.

Parker, E.N.: 1963, *Interplanetary Dynamical Processes*, Interscience, NewYork.

Parkhomov, V.A., Riazantseva, M.O., and Zastenker, G.N.: 2005, Local amplification of auroral electrojet as response to sharp solar wind dynamic pressure change on June 26, 1998. *Planet. Space Sci*. **53**, 265-274.

Phan, T.D., Gosling, J.T., and Davis, M.S.: 2009, Prevalence of extended reconnection X-lines in the solar wind at 1 AU. *Geophys. Res. Lett*. **36,** L09108.

Qin G. and Li G.: 2008, Effect of flux tubes in the solar wind on the diffusion of energetic particles. *Astrophys. J*. **682**, L129–L132.

Riazantseva, M.O., Dalin, P.A., and Zastenker, G.N.: 2002, Statistical analysis of fast and large impulses of solar wind ion flux (density) as measured by *Interball-1*. *Soln.-zemn. fizika* **2**, 89-92.

Riazantseva, M.O., Dalin, P.A., Zastenker, G.N., Parhomov, V.A., Eselevich, V.G., Eselevich, M.V., and Richardson, J.: 2003a, Properties of Sharp and Large Changes in the Solar Wind Ion Flux (Density). *Cosmic Res.* **41**, 395–404.

Riazantseva, M.O., Dalin, P.A., Zastenker, G.N., and Richardson, J.: 2003b, Orientation of sharp fronts in the solar wind plasma, *Cosmic Res.* **41**, 405–416.

Riazantseva, M.O., Khabarova, O.V., and Zastenker, G.N.: 2005, Sharp boundaries of solar wind plasma structures and an analysis of their pressure balance. *Cosmic Res.* **43**, 157-164.

Riazantseva, M.O., Khabarova, O.V., Zastenker, G.N., and Richardson, J.D.: 2007, Sharp boundaries of the solar wind plasma structures and their relationship to the solar wind turbulence. *Adv. Space Res*. **40**, 1802-1806.

Riazantseva, M.O, Zastenker, G.N., Richardson, J.D., and Eiges, P.E.: 2005, Sharp boundaries of small- and middle-scale solar wind structures, *J. Geophys. Res.* **110**, A12110.

Roberts, D.A., Keiter P.A., and Goldstein M.L.: 2005, Origin and dynamics of the heliospheric streamer belt and current sheet. *J. Geophys. Res*. **110**, A06102.

Romanova, N., Pilipenko, V., Crosby, N., and Khabarova, O.: 2007, ULF wave index and its possible applications in space physics. *Bulgarian J.Phys*. **34**, 136-148 http://www.bjp-bg.com/papers/bjp2007_2_136-148.pdf .

Safrankova, J., Zastenker, G.N., Nemecek, Z., *et al*.: 1997, Small scale observations of magnetopause motion: preliminary results of the INTERBALL project, *Ann. Geophys*., **15**, 562-569.

Schwenn, R.: 1990, Large-scale structure of the interplanetary medium. In: Schewenn R. and Marsch E. (eds.) *Physics of the Inner Heliosphere I*, Springer-Verlag, XI Berlin, Germany, 99–181.

Smith, E. J.: 2001, The heliospheric current sheet. *J. Geophys. Res*., **106**, 15819-15831.

Steiger, von R., Schwadron, N., Fisk, L., Geiss, J., Gloeckler, G., Hefti, S., *et al*.: 2000, Composition of quasi-stationary solar wind flows from Ulysses/Solar Wind Ion Composition Spectrometer. *J. Geophys. Res*. **105**, 27217-27238.

Svalgaard, L., Wilcox, J.M., Scherrer P.H., and Howard R.: 1975, The Sun's magnetic sector structure, *Solar Phys.* **45**, 83-91.

Svalgaard, L.: 1973, Geomagnetic responses to the solar wind and solar activity. NASA SUIPR report no.555, http://www.leif.org/research/Geomagnetic-Response-to-Solar-Wind.pdf

Svalgaard, L.: 1975, On the use of Godhavn H component as an indictor of the interplanetary sector polarity. *J. Geophys. Res*. **80**, 2717-2722.

Velli, M. and Grappin, R.: 1993, Properties of the solar wind. *Adv. Space Res.* **13**, 49-58.

Wang, Y.-M.; Sheeley, N.R.Jr.: 1990, Solar wind speed and coronal flux-tube expansion. *Astrophys. J.* **355**, 726-732.

Wang, Y.-M: 1993, Flux-tube divergence, coronal heating, and the solar wind. *Astrophys. J. Letters* 410, L123-L126.

Watari, S., and Watanabe, T.: 2006, Sector boundary crossings and geomagnetic activities, *Adv. in Geosciences* **2 ST**, 135-142.

Wilcox, J.M. and Ness, N.F.: 1965, Quasi-stationary corotating structure in the interplanetary medium. *J. Geophys. Res.* **70,** 5793-5805.

**Figures**

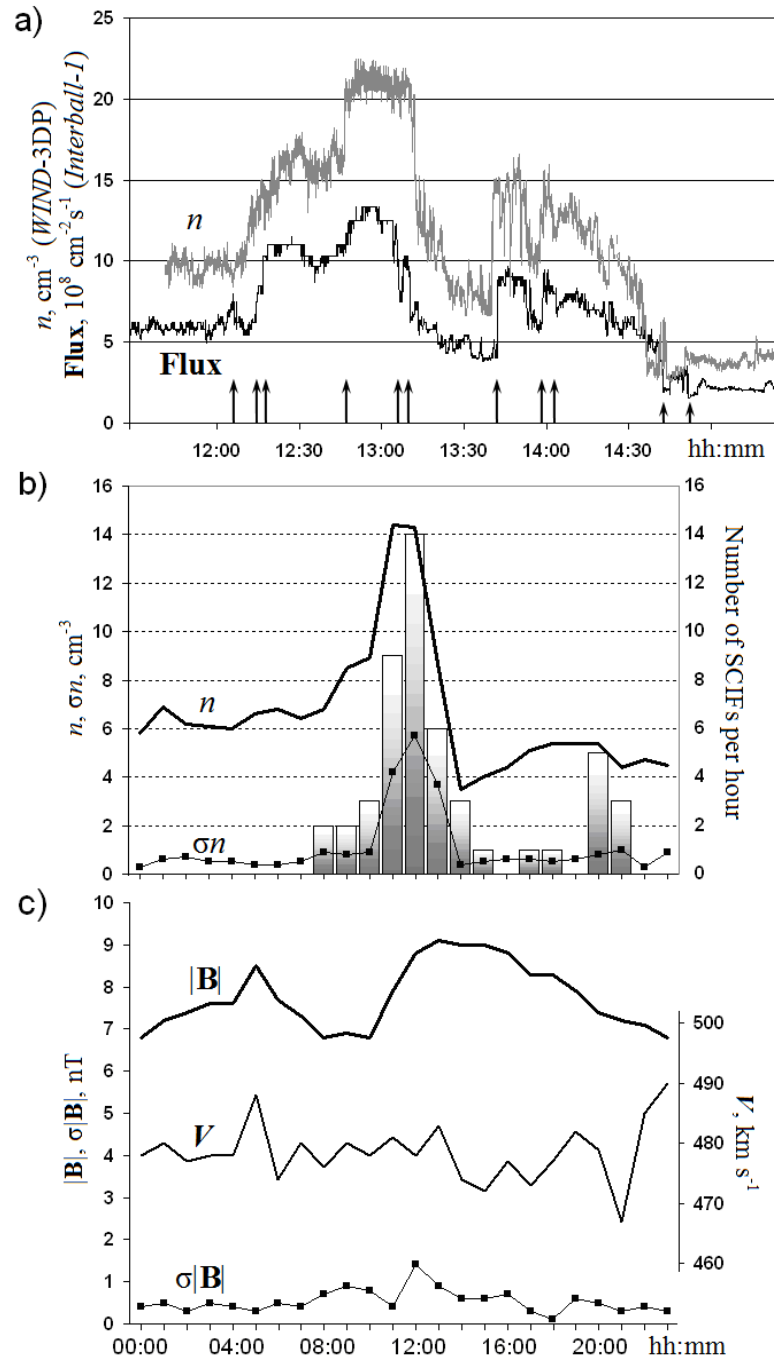

**Figure 1** Typical case of the observation of sharp ion flux and density increases/decreases on 26 April, 1998. **(a)** Solar wind density *n* (*WIND*) and ion flux **Flux** (*Interball-1*) high-resolution time series. Onsets of SCIFs with amplitudes $\geq 2\times 10^8$ cm$^{-2}$ s$^{-1}$ are pointed with arrows in **Flux**. **(b)** Vertical boxes show number of SCIFs with amplitude $\geq 0.5\times 10^8$ cm$^{-2}$ s$^{-1}$ per hour. Time series of hourly OMNI2 data ***n***, interplanetary magnetic field averaged magnitude |**B**|, solar wind speed *V*, and standard deviations from mean σ*n*, and σ|**B**| for April 26, 1998 are given in **(b)** and **(c)**.

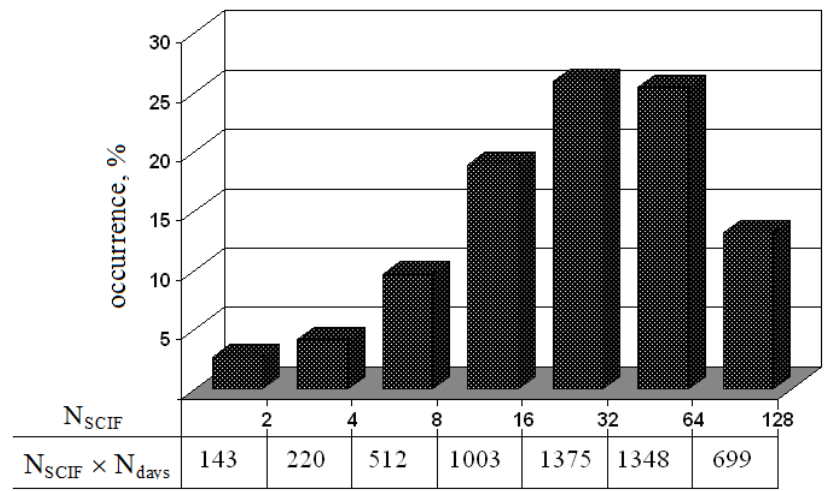

**Figure 2** Distribution of 5300 SCIFs (amplitude ≥ $2\times10^8$ cm$^{-2}$ s$^{-1}$) as a percentage of the whole number of events for the period 1996–2000.

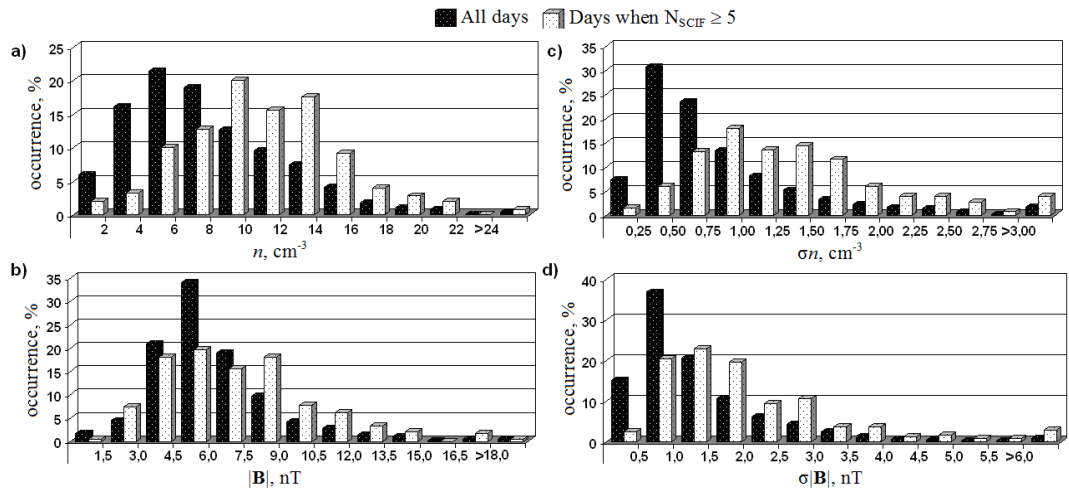

**Figure 3** Distribution of daily averaged solar wind parameters: density **(a)**, averaged interplanetary magnetic field magnitude **(b)**, and their standard deviations **(c)** and **(d)** for the days when the number of SCIFs per day ($N_{SCIF}$) exceeded five (white histograms) in comparison with distributions of the same parameters from *WIND* data (black histograms) for the period 1996–2000.

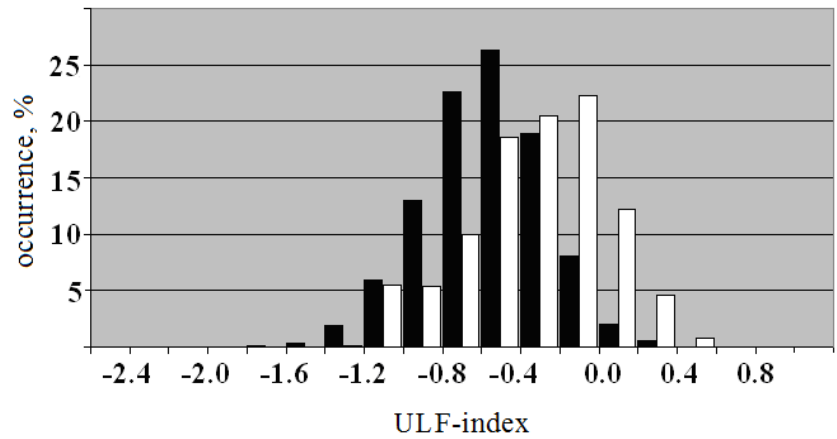

**Figure 4** Histograms of the distribution of the daily values of the interplanetary ULF-wave index for days of high SCIFs number (white histogram) and for the whole period of measurements 1996–2000 (black histogram).

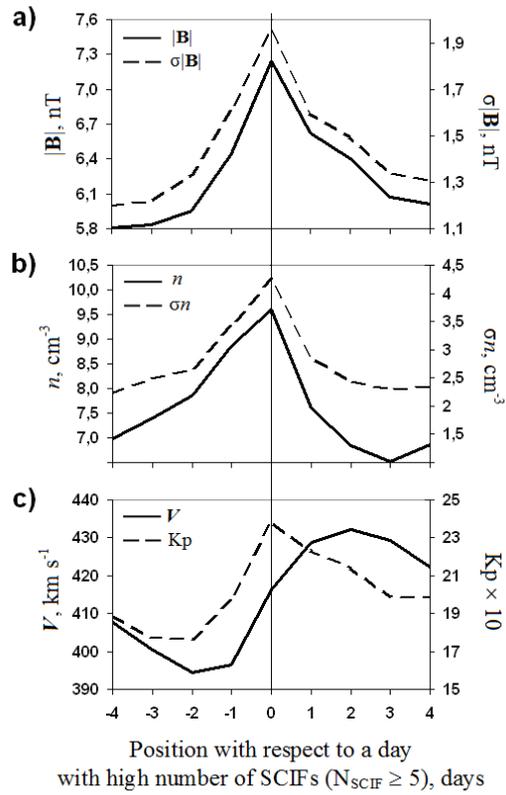

**Figure 5** Superposed epoch analysis results for solar wind parameters around the days of high SCIF number (N$_{SCIF}$ ≥ 5), 264 events. Daily values of solar wind IMF averaged magnitude |**B**|, density *n*, speed *V*, standard deviations σ|**B**|, σ*n* and Kp index of geomagnetic activity in a time range of ±4 days around day zero.

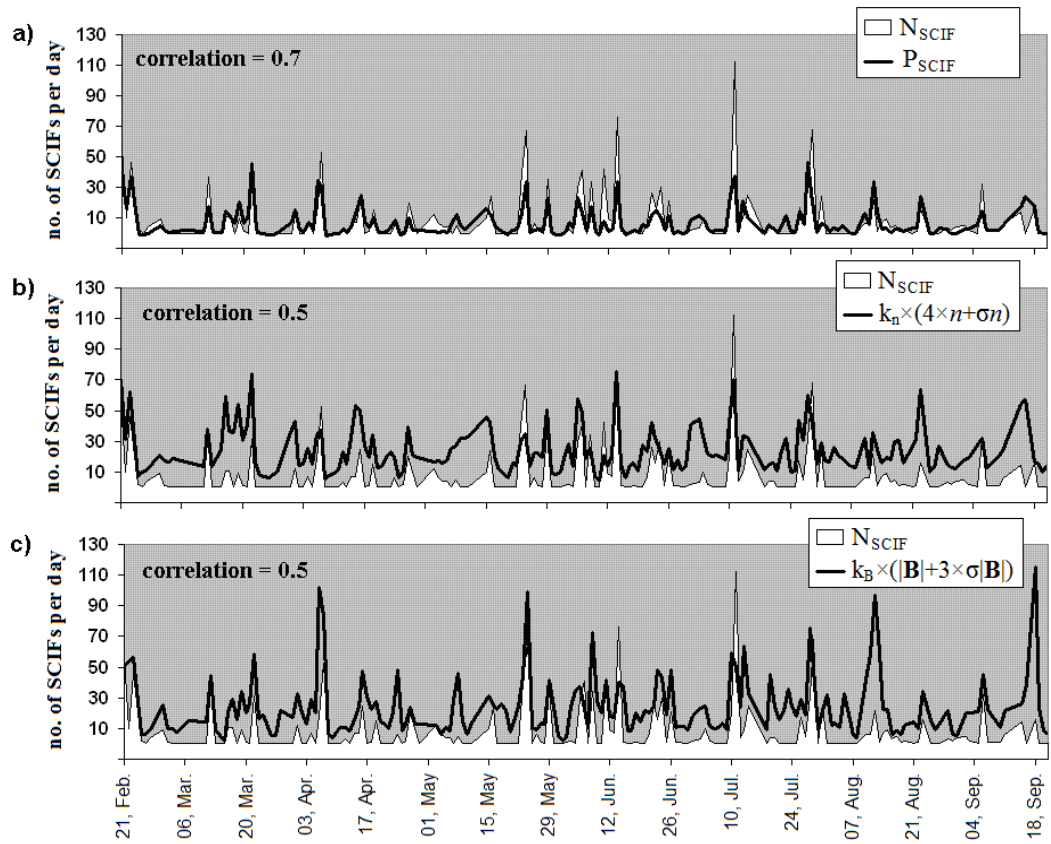

**Figure 6** Simulation of the SCIF number per day. $N_{SCIF}$ (white filled curve) is the observed number of sharp ion flux changes per day by *Interball-1* in 2000. **(a)** $P_{SCIF}$ (black curve) - modelling parameter. **(b)** and **(c)** - plasma and IMF multipliers of the modelling parameter $P_{SCIF}$.

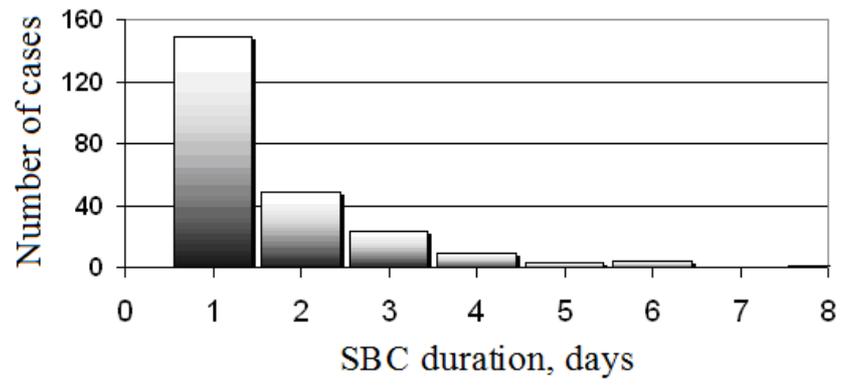

**Figure 7** Durations of sector boundary crossings for the period 1994–2000 according to the ISTP Solar Wind Catalog Candidate Events.

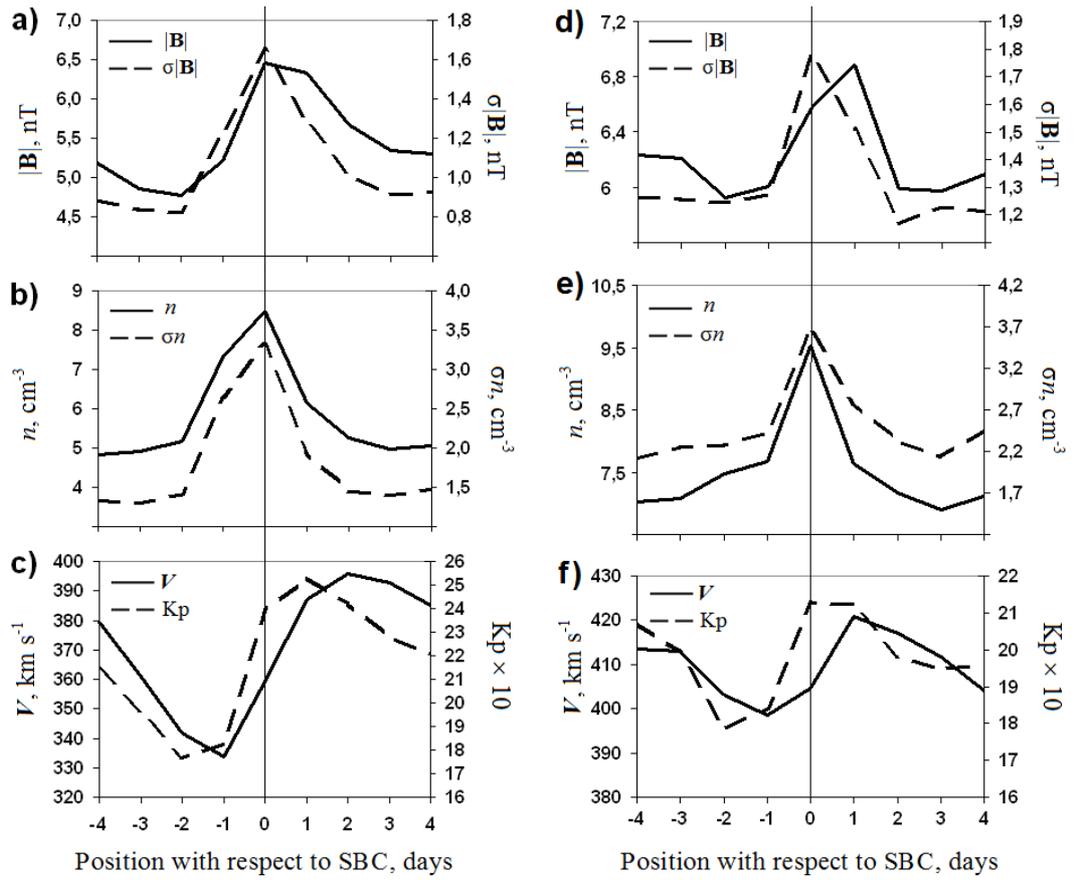

**Figure 8** Same as Figure 5, but day zero corresponds to the day of sector boundary crossing. Behaviour of parameters **(a–c)** for 1300 events from January, 1964 to April, 2010 from the SBC list by Leif Svalgaard, and **(d–f)** for 149 events of one-day sector boundaries crossings from the ISTP Solar Wind Catalogue of Candidate Events for the period 1994–2000.

**Figure legends**

**Figure 1** Typical case of the observation of sharp ion flux and density increases/decreases on 26 April, 1998. **(a)** Solar wind density *n* (*WIND*) and ion flux **Flux** (*Interball-1*) high-resolution time series. Onsets of SCIFs with amplitudes $\geq 2\times 10^8$ cm$^{-2}$ s$^{-1}$ are pointed with arrows in **Flux**. **(b)** Vertical boxes show number of SCIFs with amplitude $\geq 0.5\times 10^8$ cm$^{-2}$ s$^{-1}$ per hour. Time series of hourly OMNI2 data ***n***, interplanetary magnetic field averaged magnitude |**B**|, solar wind speed *V*, and standard deviations from mean σ*n*, and σ|**B**| for April 26, 1998 are given in **(b)** and **(c)**.

**Figure 2** Distribution of 5300 SCIFs (amplitude $\geq 2\times 10^8$ cm$^{-2}$ s$^{-1}$) as a percentage of the whole number of events for the period 1996–2000.

**Figure 3** Distribution of daily averaged solar wind parameters: density **(a)**, averaged interplanetary magnetic field magnitude **(b)**, and their standard deviations **(c)** and **(d)** for the days when the number of SCIFs per day (N$_{SCIF}$) exceeded five (white histograms) in comparison with distributions of the same parameters from *WIND* data (black histograms) for the period 1996–2000.

**Figure 4** Histograms of the distribution of the daily values of the interplanetary ULF-wave index for days of high SCIFs number (white histogram) and for the whole period of measurements 1996–2000 (black histogram).

**Figure 5** Superposed epoch analysis results for solar wind parameters around the days of high SCIF number (N$_{SCIF} \geq 5$), 264 events. Daily values of solar wind IMF averaged magnitude |**B**|, density *n*, speed *V*, standard deviations σ|**B**|, σ*n* and Kp index of geomagnetic activity in a time range of ±4 days around day zero.

**Figure 6** Simulation of the SCIF number per day. N$_{SCIF}$ (white filled curve) is the observed number of sharp ion flux changes per day by *Interball-1* in 2000. **(a)** P$_{SCIF}$ (black curve) - modelling parameter. **(b)** and **(c)** - plasma and IMF multipliers of the modelling parameter P$_{SCIF}$.

**Figure 7** Durations of sector boundary crossings for the period 1994–2000 according to the ISTP Solar Wind Catalog Candidate Events.

**Figure 8** Same as Figure 5, but day zero corresponds to the day of sector boundary crossing. Behaviour of parameters **(a–c)** for 1300 events from January, 1964 to April, 2010 from the SBC list by Leif Svalgaard, and **(d–f)** for 149 events of one-day sector boundaries crossings from the ISTP Solar Wind Catalogue of Candidate Events for the period 1994–2000.

**Tables**

**Table 1** Mean value, median, standard deviation and skewness for the solar wind parameters in Figure 3.

|  | **Mean** | **Median** | **Std.Dev.** | **Skewness** | **Valid N** |
|---|---|---|---|---|---|
| $n_{all}$, cm$^{-3}$ | 7.5 | 6.6 | 4.3 | 1.0 | 1557 |
| $n_{scif}$, cm$^{-3}$ | 10.5 | 10.3 | 4.4 | 0.4 | 250 |
| $\sigma n_{all}$, cm$^{-3}$ | 0.8 | 0.6 | 0.7 | 3.2 | 1555 |
| $\sigma n_{scif}$, cm$^{-3}$ | 1.4 | 1.2 | 0.9 | 2.8 | 250 |
| $|\mathbf{B}_{all}|$, nT | 6.0 | 5.5 | 2.6 | 1.7 | 1546 |
| $|\mathbf{B}_{scif}|$, nT | 6.9 | 6.4 | 3.2 | 1.0 | 245 |
| $\sigma|\mathbf{B}_{all}|$, nT | 1.3 | 1.0 | 1.2 | 4.2 | 1546 |
| $\sigma|\mathbf{B}_{scif}|$, nT | 2.0 | 1.6 | 1.4 | 2.1 | 245 |

**Table 2** Mean values, 95% confidence interval (conf. int.), and standard deviations in the extreme points (Std.Dev.extr.) for the solar wind parameters and the Kp index of geomagnetic activity in Figure 5.

|  | $|\mathbf{B}|$ | $\sigma|\mathbf{B}|$ | $n$ | $\sigma n$ | $V$ | Kp |
|---|---|---|---|---|---|---|
| **Mean** | 6.23 | 1.42 | 7.50 | 2.70 | 414.4 | 20.04 |
| **95% conf. int.** | 0.41 | 0.18 | 0.49 | 0.37 | 10.8 | 1.46 |
| **Std.Dev.extr.** | 3.36 | 1.46 | 4.10 | 3.1 | 89.9 | 12.10 |

**Table 3** Correlation coefficients of daily averaged solar wind parameters with SCIF number per day $N_{SCIF}$ (low correlation)

|  | $N_{SCIF}$ |
|---|---|
| $V$ | 0.07 |
| $\sigma V$ | 0.22 |
| **E**-field ($-V \times \mathbf{Bz}$) | 0.02 |
| Plasma $\beta$ | 0.05 |
| Mach number | 0.05 |

**Table 4** Correlation coefficients of daily averaged solar wind parameters with SCIF number per day $N_{SCIF}$ (moderate correlation)

|  | $N_{SCIF}$ |
|---|---|
| $n$ | 0.5 |
| $\sigma n$ | 0.3 |
| $|\mathbf{B}|$ | 0.4 |
| $\sigma|\mathbf{B}|$ | 0.4 |
| $4 \times n + \sigma n$ | 0.5 |
| $|\mathbf{B}| + 3 \times \sigma|\mathbf{B}|$ | 0.5 |

**Table 5** Mean values, 95% confidence interval (conf. int.), and standard deviation in the maximum points (Std.Dev.max.) of the solar wind parameters as well as Kp index of geomagnetic activity (Figure 8a, b and c)

|                | \|B\| | σ\|B\| | *n* | σ*n* | *V* | Kp |
|---|---|---|---|---|---|---|
| **Mean**           | 5.44 | 1.05 | 5.67 | 1.76 | 373.4 | 21.84 |
| **95% conf. int.** | 0.21 | 0.08 | 0.34 | 0.19 | 10.5  | 0.64  |
| **Std.Dev.max.**   | 3.93 | 1.55 | 6.31 | 3.45 | 193   | 11.8  |

**Table 6** Same as Table 5, but for Figure 8d, e and f

|                | \|B\| | σ\|B\| | *n* | σ*n* | *V* | Kp |
|---|---|---|---|---|---|---|
| **Mean**           | 6.25 | 1.34 | 7.42 | 2.44 | 410.0 | 19.92 |
| **95% conf. int.** | 0.35 | 0.19 | 0.90 | 0.59 | 13.6  | 1.69  |
| **Std.Dev.max.**   | 2.15 | 1.19 | 5.63 | 3.68 | 84.6  | 10.51 |